\documentclass[referee]{aa}  

\usepackage{graphicx}
\usepackage{txfonts}
\usepackage{graphics,graphicx,rotating,amsmath,color}

\newcommand{\bth}{{\boldsymbol \theta}}
\newcommand{\bde}{{\boldsymbol \delta}}
\newcommand{\bep}{{\boldsymbol \epsilon}}

\newcommand{\mbe}{{\mbox{\boldmath$e$}}}

\newcommand{\mbg}{{\mbox{\boldmath$g$}}}
\newcommand{\mbk}{{\mbox{\boldmath$k$}}}
\newcommand{\cZ}{{\cal Z}}

\begin{document}

\title{The ERA Method with Idealizing PSF for Precise Weak Gravitational Lensing Shear Analysis}
\author{Yuki Okura\inst{1} \and  Toshifumi Futamase\inst{2}}
\institute{RIKEN, yuki.okura.2014@gmail.com \and 
Tohoku University, tof@astr.tohoku.ac.jp
}

\abstract{
We generalize ERA method of PSF correction for more realistic situations. The method   re-smears the observed galaxy image(galaxy image smeared by PSF) and PSF image by 
an appropriate function called Re-Smearing Function(RSF) to make new images which have the same ellipticity with the lensed (before smeared by PSF) galaxy image. It has been shown that the method avoids a systematic error arising from an approximation in the usual PSF correction in moment method such as KSB for simple 
PSF shape.  By adopting an idealized PSF  we generalize ERA method applicable for arbitrary PSF. This is confirmed with simulated complex PSF shapes.
We also consider the effect of pixel noise and found that the effect causes systematic overestimation. 
}


\maketitle

\section{Introduction}
It is now widely recognized that weak gravitational lensing is an unique and powerful tool to obtain mass distribution in the universe. Coherent deformation of the shapes of background galaxies carries not only the information of intervening mass distribution but also the  cosmological background geometry and thus the cosmological parameters(Mellier 1999, Schneider 2006, Munshi et al. 2008). 

In fact weak lensing studies have revealed the averaged mass profile for galaxy cluster
(Okabe et al. 2013,  Umetsu et al. 2014) and detected the cosmic shear that is weak lensing by large scale structure is expected to be useful for studying the property of dark energy.   However the signal of cosmic shear is very weak and difficult to get useful constraint on the dark energy.   
Currently, several surveys are just started and planned to measure the cosmic shear accurately enough to constrain the dark energy property, such as Hyper Suprime-Cam on Subaru (http://www.naoj.org/Projects/HSC/HSCProject.html),  EUCLID (http://sci.esa.int/euclid) and LSST (http://www.lsst.org). 
Since the signal of cosmic shear is very small, these surveys plan to observe a huge number large of background galaxies to reduce statistical error. 
This means that any systematic errors in the lensing analysis must be controlled to be 
smaller than the statistic error, roughly saying $1 \% \sim 0.1 \%$ error is required for the systematic error.  
For this purpose there have been many methods(Bernstein \& Jarvis 2002; Refregier 2003; Kuijken et al. 2006; Miller et al. 2007; Kitching et al. 2008; Melchior 2011) have been developed and tested with simulation(Heymans et al 2006, Massey et al 2007, Bridle et al 2010 and Kitching et al 2012) .  Although there have been a great progress, it seems that no fully satisfying method is available yet. 

One of the systematic error comes from smearing effect by atmospheric turbulence and imperfect optics. This effect is described by point spread function(PSF) and we need to correct the effect very accurately in order to study the dark energy property . 
Previous approaches of PSF correction adopted some sort of  approximation for the form of PSF which prevents from an accurate correction in some cases.
Recently, we have proposed a new approximation free method of PSF correction called 
ERA method  (ERA1:Okura and Futamase 2014, ERA2:Okura and Futamase 2015) based on 
E-HOLICs method(Okura and Futamase 2011, Okura and Futamase 2012, Okura and Futamase 2013)  which is a generalization of KSB method(Kaiser at al. 1995) and 
uses an elliptical weight function to avoid  expansion of weight function in measuring ellipticity. 
The method makes use of the artificial image constructed by re-smearing the observed image to have the same ellipticity with the lensed image. We have confirmed by numerical simulation that the method is free from systematic error, but is restricted to the case that PSF has a simple form.  In this paper we generalize ERA method in more realistic shape of PSF and show that there is no systematic error by numerical simulation.  

The plan of this paper is as follows.
In section 2, we explain about definitions and notations used in this paper.
In section 3, we explain ERA method and then generalize it for more realistic shape 
of PSF by idealizing the original PSF shape. 
In section 4, we show the results of simulation tests for this method with complex PSF shape
and with pixel noise on galaxy image and PSF. 
Finally we summarize our method and give some discussion  in section 5. 

\section{Definitions and Notations}
In this section, we explain the original definitions and notations used in this paper for reader's convenience. More details may be found in ERA1 and ERA2.
The general introduction of weak gravitational lensing can be found, for example,  in Bartelmann M. \& Schneider P., 2001.

\subsection{Zero Plane}
\begin{figure*}[htbp]
\centering
\resizebox{0.75\hsize}{!}{\includegraphics{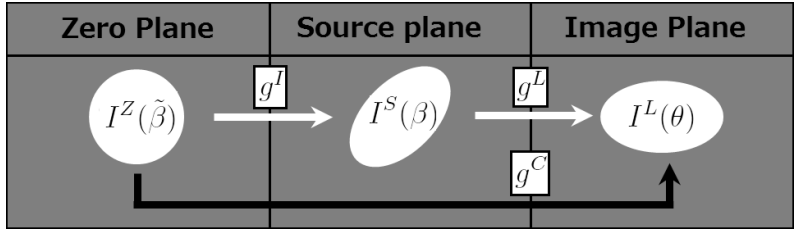}}
\caption{
\label{fig:ERA_system0}
The relation between Zero, Source and Image plane.
}
\end{figure*}

We use the concept  of "Zero Plane"  which treats the intrinsic ellipticity of source image $I^S$ comes from an imaginary distortion from circular image in zero plane $I^Z$. 
The distortion is called as intrinsic shear $\mbg^I$. This means that the lensed image $I^L$ is distorted by shear $\mbg^C$ which is described by combination of intrinsic shear $\mbg^I$ and lensing shear $\mbg^L$ as follows
\begin{eqnarray}
\label{eq:Cshear}
\mbg^C \equiv \frac{\mbg^I+\mbg^L}{1+\mbg^I\mbg^{L*}},
\end{eqnarray}
Fig.\ref{fig:ERA_system0} is an illustration about the relation between Zero, Source and Image plane. 
Since $\mbg^I$ has random orientation, the average of the shear is 0, so we obtain
\begin{eqnarray}
\label{eq:avegC}
\left< \frac{\mbg^C-\mbg^L}{1-\mbg^C\mbg^{L*}} \right>=\left< \mbg^I \right>=0,
\end{eqnarray}
Thus we obtain the weak lensing shear $\mbg^L$ from average of $\mbg^C$ as satisfying eq.\ref{eq:avegC}.

\subsection{Weak lensing shear and Ellipticities}

Now we introduce two kind of ellipticities used in ERA method. 
 
The complex image moments are  measured as 
\begin{eqnarray}
\label{eq:CMOM}
\cZ^N_M(I,\bep_W)&\equiv&\int d^2\theta \bth^N_M I(\bth) W(\bth,\bep_W)\\
\bth^N_M&\equiv&\bth^{\frac{N+M}{2}}\bth^{*\frac{N-M}{2}},
\end{eqnarray}
where $\bth^N_M$ is the higher order complex displacement from the centroid of the image and N means the order of $\bth$ and M means the spin number, and $W(\bth,\bep_W)$ is the weight function with ellipticity $\bep_W$. 

We also make use of the non-dimensional moment defined as 
\begin{eqnarray}
\label{eq:CMOM02}
\cZ^0_2(I,\bep_W)&\equiv&\int d^2\theta \frac{\bth^2_2}{\bth^2_0} I(\bth) W(\bth,\bep_W)
\end{eqnarray}

The above moments naturally lead to the following ellipticities. 
\begin{eqnarray}
\label{eq:2ndE}
\bep_{2nd}&\equiv&\frac{\cZ^2_2(I,\bep_W)}{\cZ^2_0(I,\bep_W)}\\
\label{eq:0thE}
\mbe_{0th}&\equiv&\frac{\cZ^0_2(I,\bep_W)}{\cZ^0_0(I,\bep_W)},
\end{eqnarray}
where the weight function should be same as measured ellipticity, so $\bep_W=\bep_{2nd}$ for eq.\ref{eq:2ndE} and $\bep_W=\bep_{0th}\equiv2\mbe_{0th}/(1+|\mbe_{0th}|^2)$ for eq.\ref{eq:0thE}. $\bep_{2nd}$ is the usual ellipticity and we call $\mbe_{0th}$ as the 0th ellipticity. In generally, the 0th-ellipticity has higher signal to noise ratio than 2nd-ellipticity, because 0th-ellipticity is measured from central region of images(see ERA2). 

These two ellipticities are related to the lensing shear as follows
\begin{eqnarray}
\label{eq:2ndEshear}
\bep&=&\bde\equiv\frac{2\mbg}{1+|\mbg|^2}\\
\label{eq:0thEshear}
\mbe&=&\mbg \hspace{56pt} |\mbg|<1\nonumber\\
        &=&\frac{1}{\mbg^*} \hspace{50pt} |\mbg|>1,
\end{eqnarray}
where $\bde$ is lensing complex distortion. 

Let us summarize steps to measure lensing reduced shear $\mbg^L$ is as follows. 
1) Measure the 2nd and/or 0th ellipticity calculated as eq.\ref{eq:2ndE} and/or eq.\ref{eq:0thE} using image moments measured as eq.\ref{eq:CMOM} and/or eq.\ref{eq:CMOM02}, 
2) Then calculate the combined reduced shear $\mbg$ from the ellipticities as eq.\ref{eq:2ndEshear} and/or eq.\ref{eq:0thEshear}.
3) Finally  obtain the lensing reduced shear from combined shear by averaging as satisfying eq.\ref{eq:avegC}.

\section{The Basis of ERA Method and Idealizing PSF Shape}
In this section, we explain the outlines of the PSF correction in ERA method. More details ,can be seen in ERA1.

\subsection{The Basis of PSF correction in the ERA Method}
In real analysis, we cannot observe the galaxy image (GAL) $I^{GAL}$ directly 
because of the smearing due to various effects such as atmospheric turbulence and imperfect optics. What we observe is the smeared image (SMD) $I^{SMD}$. These effects 
are supposed to be described in Fourier space as follows.
\begin{eqnarray}
\label{eq:SMDGAL_F}
\hat I^{SMD}(\mbk)=\hat I^{GAL}(\mbk)\hat I^{PSF}(\mbk),
\end{eqnarray}
where hat means Fourier transformed function of from the function in the two-dimensional angular plane. The function expressing the smearing is called as Point Spread Function(PSF) $I^{PSF}$, and PSF can be measured from star image because star image is a point source before the smearing, so the brightness distribution of star is PSF at the position of  stars. Therefore we can obtain ellipticities of $I^{GAL}$ from $I^{SMD}$ with PSF correction using $I^{PSF}$. 
However, some PSF correction methods introduce a systematic error due to insufficient correction.

We have developed a new PSF correction method called ERA method. The idea  is to re-smear SMD and PSF by a Re-Smearing Function (RSF) $I^{RSF}$ as follows.
\begin{eqnarray}
\label{eq:ReSMDGPSF_F}
\hat I^{RePSF}(\mbk)&=&\hat I^{PSF}(\mbk)\hat I^{RSF}(\mbk)\\
\label{eq:ReSMDGAL_F}
\hat I^{ReSMD}(\mbk)&=&\hat I^{SMD}(\mbk)\hat I^{RSF}(\mbk)=\hat I^{GAL}(\mbk)\hat I^{PSF}(\mbk)\hat I^{RSF}(\mbk)\nonumber\\
&=&\hat I^{GAL}(\mbk)\hat I^{RePSF}(\mbk),
\end{eqnarray}
These formulas mean that RSF makes new images; Re-smeared PSF (RePSF) $I^{RePSF}$ and Re-smeared galaxy (ReSMD) $I^{ReSMD}$. 
Since ReSMD can be expressed as the convolution of GAL and RePSF, 
we can choose PSF in an arbitrary shape RePSF by an appropriate RSF. 
Then if ReSMD has the ellipticity same with GAL, we can obtain the ellipticity of GAL 
from that of ReSMD, and the situation occurs if RePSF has the same ellipticity with GAL and ReSMD. 

The summary of the steps in the PSF correction are as follows. 
1) First re-smear SMD and PSF by RSF(with initial trial function), 
2) Then measure the ellipticity of ReSMD and RePSF, 
3) If the ellipticities of ReSMD and RePSF do not coincide each other, then try the step  1) and 2) with corrected RSF until they coincide.  
4) If the ellipticity of ReSMD and RePSF coincide, the ellipticity is ellipticity of GAL.

We have suggested two type of RSFs in the previous papers ERA1; 
One is to use the deconvolution and an elliptical Gaussian (Method A in ERA1) and the other is to introduce a small elliptical Gaussian for RSF(Method B in ERA1). However they both have some difficulties. Method A uses a deconvolution constant to  avoid  noise enhancement and it introduces a systematic error if we use a large value  constant. On the other hand, the method B simply smears PSF, so it does not use the  deconvolution constant, but if PSF has complex shape, the ellipticity of PSF changes by measuring parameters, so it can not define the exact ellipticity of RePSF.
\subsection{Idealizing PSF Shape}
To solve the above difficulties, we developed a new PSF correction method which make an idealized PSF from PSF with arbitrary shape as follows.

1) First measure PSF 
2) consider RePSF with some ideal profile with an ellipticity $\bep_{RePSF}$ 
3) find RSF in Fourier space as follows
\begin{eqnarray}
\label{eq:RSF_F}
\hat I^{RSF}(\mbk)&=&\frac{\hat I^{RePSF}(\mbk)}{\hat I^{PSF}(\mbk)}.
\end{eqnarray}
However, eq. \ref{eq:RSF_F} has a problem if $\hat I^{RSF}(\mbk)$ has large value in some $\mbk$, because the large value enhances large pixel noise on galaxy image. Therefore we must find reasonable RePSF which doesn't make large value in RSF. 
This may be achieved by comparing value ratio in Fourier space. Namely, let's suppose that  the signal to noise ratio of PSF has a maximum signal at $\mbk_0$ in Fourier space, e.g. $\mbk_0=0$ for Gaussian PSF and we set the same value for RePSF at $\mbk_0$, so $\hat I^{RePSF}(\mbk_0)=\hat I^{PSF}(\mbk_0)$. Then we use RSF upper limit $\alpha$ for constrain RePSF in all $\mbk$ as follows
\begin{eqnarray}
\label{eq:RePSF_constrain}
\left | \frac{\hat I^{RePSF}(\mbk)}{\hat I^{PSF}(\mbk)} \right |<\alpha.
\end{eqnarray}
The upper limit $\alpha$ is introduced to avoid a large value in RSF and 
thus to reduce the pixel noise on GAL.
Fig. \ref{fig:ERA_SYSTEM_S1_R} is an example of this re-smearing. 
The actual value of  the upper limit depends on the strength of pixel noise,  
We study the upper limit in section \ref{sec:sim_pn} and in future works.
The only restriction for the choice of RSF is that ReSMD and RePSF have the same ellipticity, therefore there is an arbitrariness in the shape of RSF and  this is one of possibility.
The steps to find reasonable RSF is thus as follows; 1) measure the  PSF, 2) consider RePSF as an ideal elliptical function, and then  3) compare counts PSF and RePSF in Fourier space each $\mbk$.  4) If RePSF/PSF has larger value than RSF upper limit, then  in some $\mbk$ we reset RePSF to a different or more narrower profile in Fourier space and try 4) again.  5) If the ratio is smaller than the RSF upper limit in all $\mbk$, the RePSF/PSF is reasonable RSF.
The detailed forms of RSF under some situations are shown in simulation test in the next section.

One of the important points is that there is no restriction about definition of ellipticity in this PSF correction. Here we use the ellipticities defined from image moments in our simulation test, but ellipticity with any definition can be used if the lensing shear can be obtained from the ellipticity, e.g. ellipticity from coefficient of decomposed image, ellipticity measured by model fitting and so on.
\section{Simulation Test}
In this section, we test the precision of new ERA method with simulation data in 2 situations. One is an ideal situation in which the systematic error comes only from PSF correction. This test is not realistic but it shows us the potential precision of ERA method. Another is the situation with pixel noise. This test includes the investigation of the behaviour of the upper limit selection in eq.\ref{eq:RePSF_constrain}.

\subsection{simulation in ideal situation}
In this section, we show the results of simulation test for our PSF correction with complex PSF shape in ideal situation.
In this test we consider only systematic error in PSF correction. It means that  we do not consider other effects such as pixel noise, pixelization and PSF interpolation. 
In the following,  we use sufficiently large images to neglect pixelization effect, high signal to noise ratio images to neglect pixel noise effect and use the same PSF for galaxy with  star's PSF to neglect PSF interpolation effect.

We consider an elliptical Gaussian(GAL) with ellipticity $[0.3,0.0]$ and covers 50 pixel for Gaussian size (to ignore pixelization effect), four types of PSF shapes and RSF upper limit is 1.2.  
The upper limit is a temporal value because the upper limit should depend on pixel noise, so in the ideal situation ERA method we can use any values for the upper limit. 
We use $10^{-9}$ for the precision in the determination of ellipticity in the iteration.
In the simulation we use the following iteration 
to obtain PSF corrected ellipticity. Let suppose we have $\bep^{ReSMD}_{i}$ for the ellipticity 
of RePSF in $i$ th iteration and then obtained the ellipticity of ReSMD as  $\bep^{ReSMD}_{i}$. Then  we set the ellipticity of RePSF in $i+1$th iteration as
\begin{eqnarray}
\label{eq:iteration}
\bep^{RePSF}_{i+1}=\bep^{ReSMD}_{i},
\end{eqnarray}
with the initial value of the iteration as 
\begin{eqnarray}
\label{eq:iteration}
\bep^{RePSF}_{0}=\bep^{SMD}.
\end{eqnarray}

\begin{figure*}[htbp]
\centering
\resizebox{0.75\hsize}{!}{\includegraphics{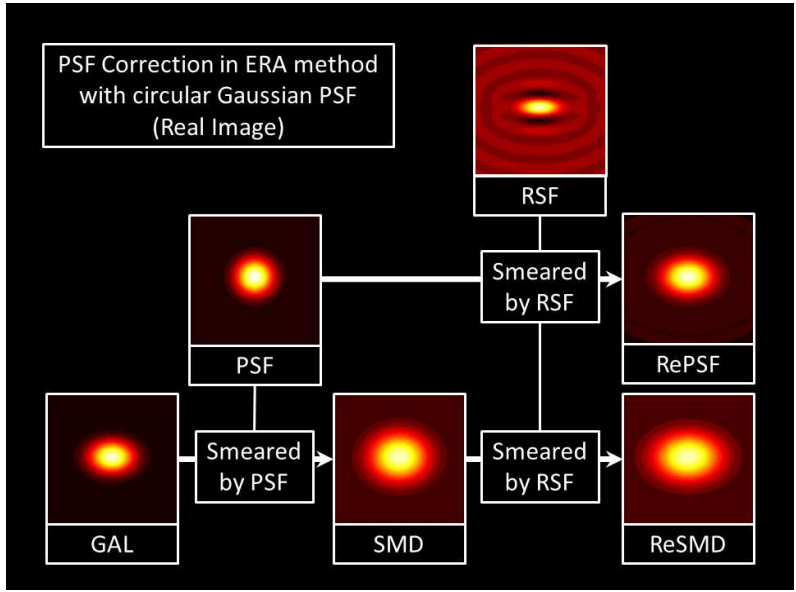}}
\caption{
\label{fig:ERA_SYSTEM_S1_R}
Illustration about simulation real images in PSF correction in ERA method with the type A.
}
\end{figure*}

\begin{figure*}[htbp]
\centering
\resizebox{0.75\hsize}{!}{\includegraphics{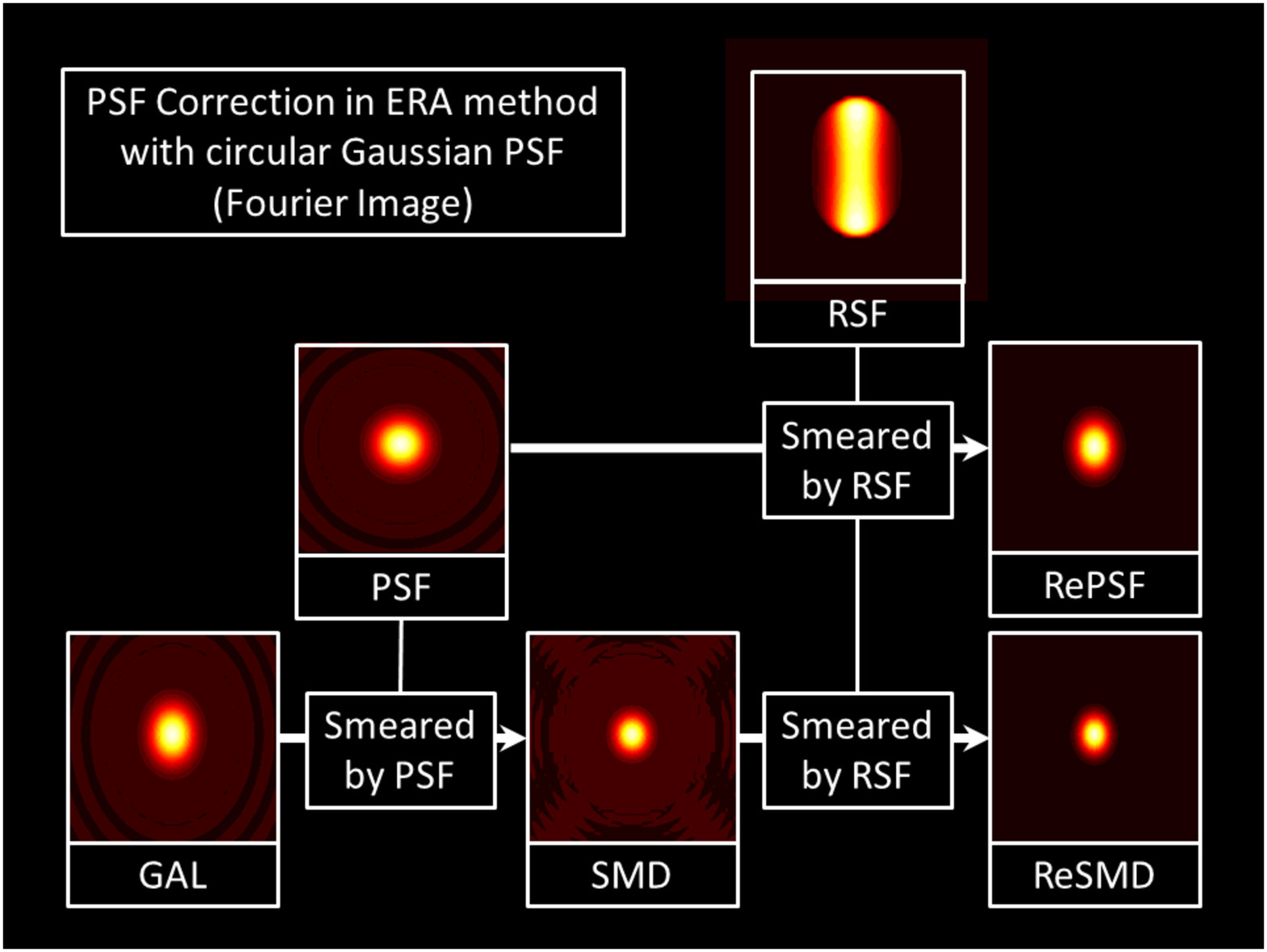}}
\caption{
\label{fig:ERA_SYSTEM_S1_F}
Illustration about simulation Fourier images in PSF correction in ERA method with the type A.
}
\end{figure*}

\begin{figure*}[htbp]
\centering
\resizebox{0.75\hsize}{!}{\includegraphics{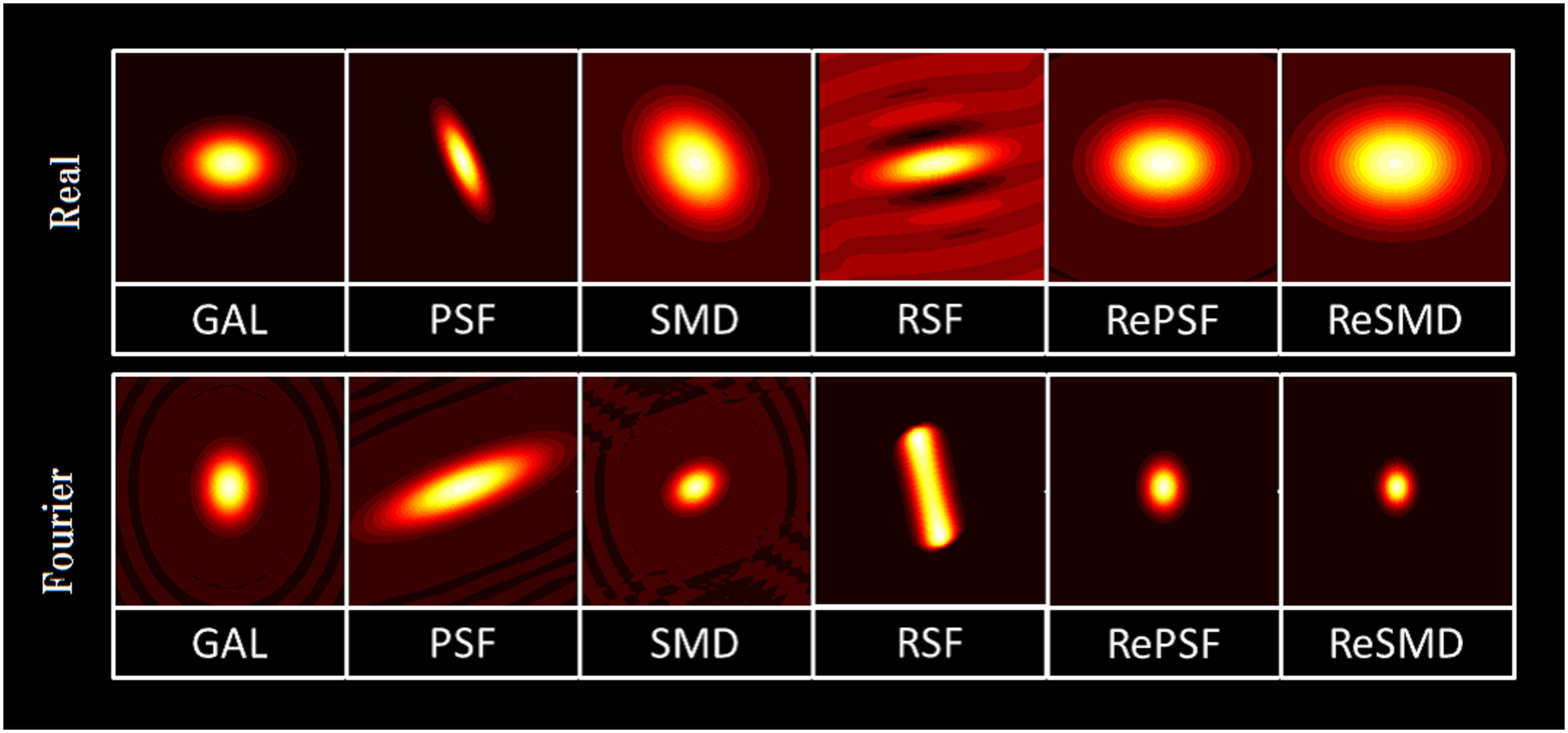}}
\caption{
\label{fig:ERA_SYSTEM_S2}
Illustration about simulation real images in PSF correction in ERA method with the type B. The labels of the images mean same as fig \ref{fig:ERA_SYSTEM_S1_R} and  fig \ref{fig:ERA_SYSTEM_S1_F}.
}
\end{figure*}

\begin{figure*}[htbp]
\centering
\resizebox{0.75\hsize}{!}{\includegraphics{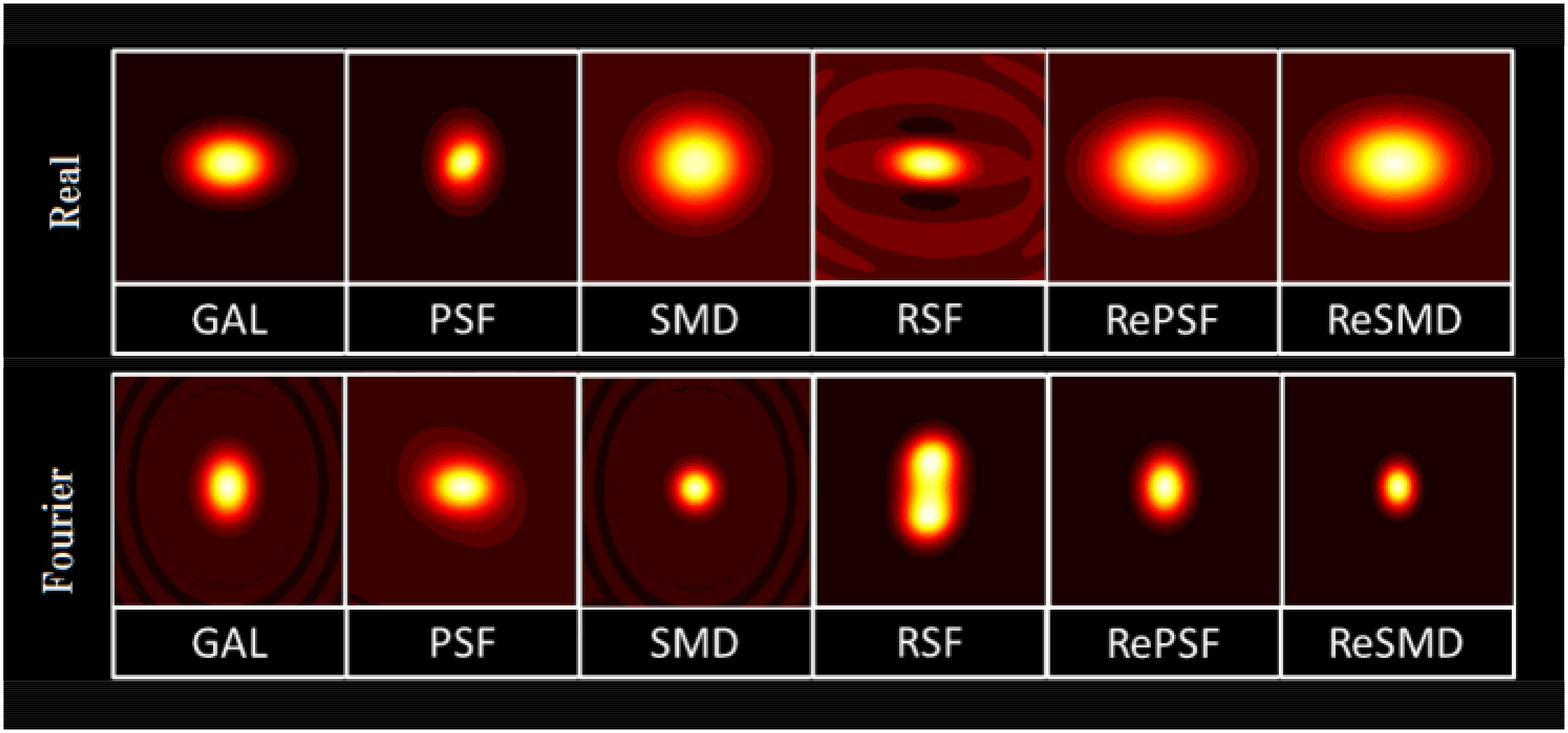}}
\caption{
\label{fig:ERA_SYSTEM_S3}
Illustration about simulation real images in PSF correction in ERA method with the type C. The labels of the images mean same as fig \ref{fig:ERA_SYSTEM_S1_R} and  fig \ref{fig:ERA_SYSTEM_S1_F}.
}
\end{figure*}

\begin{figure*}[htbp]
\centering
\resizebox{0.75\hsize}{!}{\includegraphics{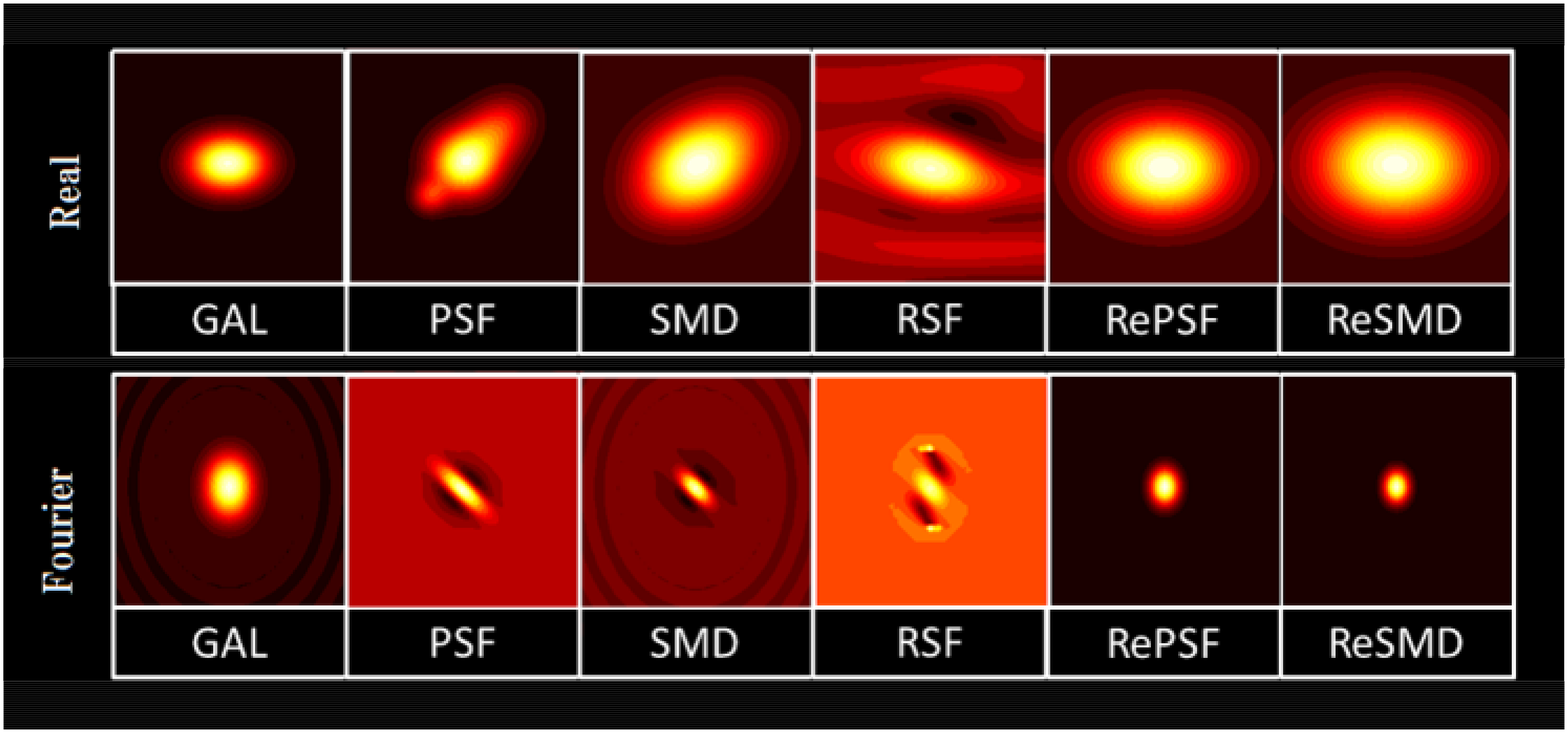}}
\caption{
\label{fig:ERA_SYSTEM_S4}
Illustration about simulation real images in PSF correction in ERA method with the type D. The labels of the images mean same as fig \ref{fig:ERA_SYSTEM_S1_R} and  fig \ref{fig:ERA_SYSTEM_S1_F}.
}
\end{figure*}

First PSF type is a circular a Gaussian PSF with the same Gaussian size as galaxy(Type A). Fig\ref{fig:ERA_SYSTEM_S1_R} and Fig\ref{fig:ERA_SYSTEM_S1_F} show GAL, PSF,  SMD, RSF, RePSF and ReSMD images in real and Fourier space(only real component), respectively.
the second type is a highly elliptical Gaussian PSF with ellipticity [-0.6,-0.6] and the Gaussian size half of the galaxy(Type B). Fig\ref{fig:ERA_SYSTEM_S2} shows images in this situation.
The third is double Gaussian PSF, one has the ellipticity [0.0,0.3] with half Gaussian size of Galaxy and the other has the ellipticity [-0.3,0.0] with the same Gaussian size of Galaxy(Type C). Fig\ref{fig:ERA_SYSTEM_S3} show images in this situation.
The fourth is three Gaussian PSF, the first has circular shape with the same Gaussian size as galaxy, the second has ellipticity [0.0,0.2] with 1.5 times large Gaussian size as galaxy and the position offset with length twice of galaxy size, the third  has circular shape with half  Gaussian size as galaxy and the position offset with length twice of galaxy size for different direction from 2nd Gaussian(Type D). 
This situation is supposed to express a pointing error. Fig\ref{fig:ERA_SYSTEM_S4} show images in this situation.
Table \ref{tbl:results_G} shows the results of the simulation test.
1st column means Type ID, and 2nd column explains PSF type. 
3rd column shows systematic error in the PSF correction by 0th ellipticity(defined by 0th order moments, ERA2), where the systematic error is defined as 
\begin{eqnarray}
\label{eq:iteration}
{\delta}\equiv\frac{\bep^{ReSMD}}{\bep^{GAL}}-1.
\end{eqnarray}
4th column shows the same with 3rd column expect that 2nd ellipticity(defined by quadrupole moments) is used.
Then, we tested again with Sersic type galaxies which has Gaussian fitted size approximately 50 pixel.
Table \ref{tbl:results_S} shows the results of the simulation test.

In the all tests, we are able to obtain the ellipticity of GAL from ellipticity of ReSMD with 
error in the range $10^{-9}$ to $10^{-5}$.
We guess that the error might be originated from the fact that we use a finite number of pixels (i.e. pixelization), 
but we was not able to test with more large images, because we cannot use infinite number of pixels for simulation.
\begin{table}
\begin{tabular}{|c|c|c|c|}\hline
Type ID & PSF type &error ratio $\times 10^{-5}$ & error ratio $\times 10^{-5}$
\\
&&2nd ellipticity& 0th ellipticity
\\\hline
A & Circular Gaussian &0.0009749 & -0.008191
\\\hline
B & High elliptical Gaussian& 0.002207 & 0.1463
\\\hline
C & Double Gaussian& 0.0003409 & 0.001661
\\\hline
D & Triple Gaussian with pointing error & -0.008504 & -1.387
\\\hline
\end{tabular}
\caption{
\label{tbl:results_G}
The PSF type and results of each situations for elliptical Gaussian galaxy.}
\end{table}\\

\begin{table}
\begin{tabular}{|c|c|c|c|}\hline
Type ID & PSF type &error ratio $\times 10^{-5}$ & error ratio $\times 10^{-5}$
\\
&&2nd ellipticity& 0th ellipticity
\\\hline
A & Circular Gaussian &0.0007257 & 3.176
\\\hline
B & High elliptical Gaussian& 0.1740 & 0.5102
\\\hline
C & Double Gaussian& -0.0003589 & 2.480
\\\hline
D & Triple Gaussian with pointing error & 0.08765 & -2.545
\\\hline
\end{tabular}
\caption{
\label{tbl:results_S}
The PSF type and results of each situations for elliptical Sersic galaxy.}
\end{table}
So we confirmed this new PSF correction with ERA method can correct PSF with enough precision.
\subsection{Simulation with pixel noise}
\label{sec:sim_pn}
\begin{figure*}[htbp]
\centering
\resizebox{0.75\hsize}{!}{\includegraphics{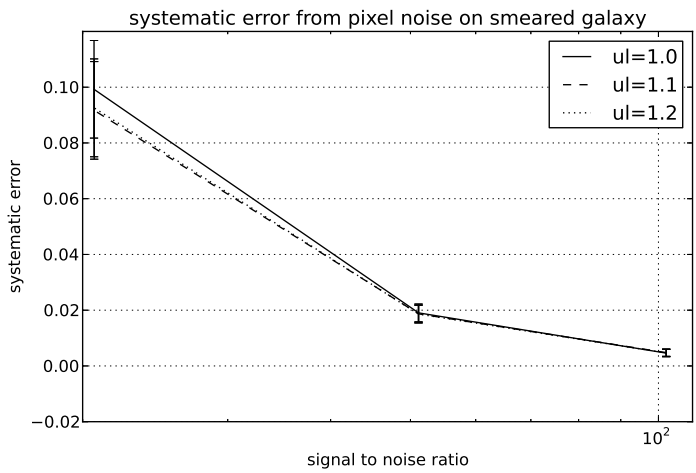}}
\caption{
\label{fig:Serror_SMD}
Systematic error for objects which have pixel noise.
}
\end{figure*}

In this section, we consider the systematic error in ERA method caused by a pixel noise. 
We consider several situations in which the galaxy is a Gaussian shape and PSF is a   circular Gaussian PSF(situation A) with 5 pixels Gaussian size.
First simulation is to study the systematic error caused by pixel noise on smeared galaxies, where the galaxies have signal to noise ratio approximately 20, 50 or 100, and PSF does not have any pixel noise. 
To see the dependence of the results on the upper limit $\alpha$ in PSF, we 
use [1.0, 1.1, 1.2] for $\alpha$. 
Fig \ref{fig:Serror_SMD} shows the result of the simulation test. It shows that the pixel noise on smeared galaxy image makes the overestimation and smaller SNR objects suffers 
more overestimation. On the other hand, no significant differences are observed with  
difference in the upper limit. 
The overestimation can be expressed by approximately as $40/({\rm SNR})^2$, and so ERA method has 1\% precision if objects have signal to noise ratio larger than approximately 60, since there is no correction for the pixel noise effect.
This means that we need pixel noise correction for precise cosmology. 

\begin{figure*}[htbp]
\centering
\resizebox{0.75\hsize}{!}{\includegraphics{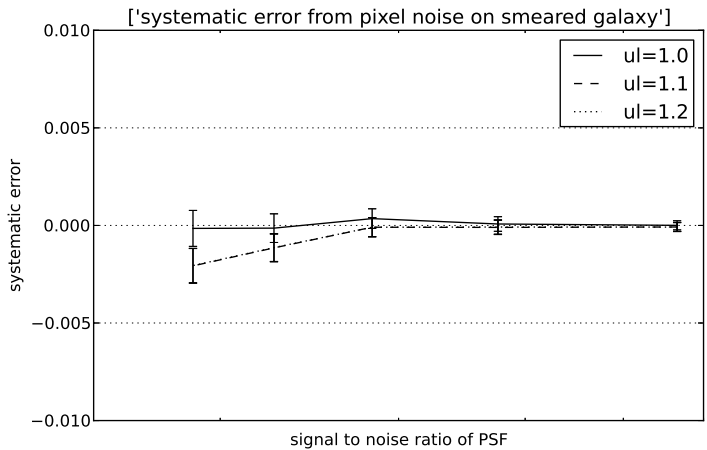}}
\caption{
\label{fig:Serror_PSF}
Systematic error for objects with PSF with pixel noise on PSF.
}
\end{figure*}

\begin{figure*}[htbp]
\centering
\resizebox{0.75\hsize}{!}{\includegraphics{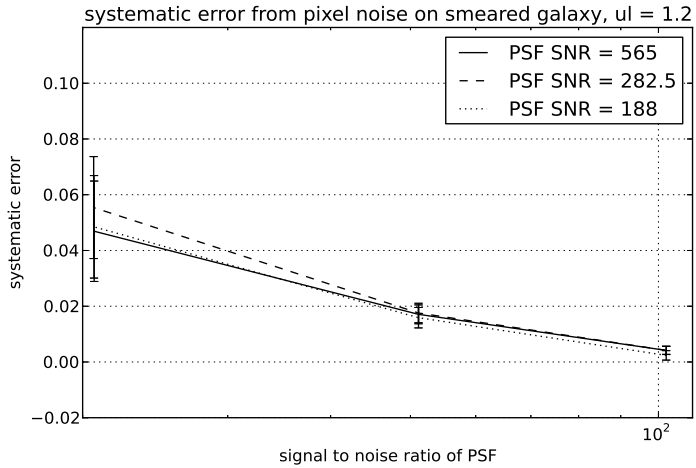}}
\caption{
\label{fig:Serror_SMDPSF}
Systematic error for objects with PSF with pixel noise on galaxy and PSF, where ul=1.2 was selected.
}
\end{figure*}

Next we consider the systematic error caused by the pixel noise on PSF, where PSF has signal to noise ratio between approximately 200 to 500, because PSF is measured from star which is much brighter than galaxy in the usual analysis. 
Fig \ref{fig:Serror_PSF} shows that the systematic error from pixel noise on PSF is under 0.005\%, and there is no significant differences in different choice of the upper limit.
Fig \ref{fig:Serror_SMDPSF} shows the systematic error from pixel noise on smeared galaxies and on PSF. Galaxies and PSF have the same signal to noise ratio as previous simulation tests. These noises cancel partly each other, but still cause net  overestimation 
and the behaviour of the figure is similar to that of first simulation tests.

The summary of these tests is that the pixel noise causes  overestimation in realistic 
situations we considered and no significant differences is found by choosing different  upper limit. Since the systematic error depends on signal to noise ratio, the pixel noise correction is urgently needed for sciences using weak lensing. 
\section{Summary}
Since the large scale cosmic shear observations are planned near future, it is urgently important to control any systematic errors as small as possible. One of serious systematic errors comes from PSF correction. Here we propose a new and approximation free PSF correction method within the framework of  ERA method which is expected to avoid systematic error associated with PSF correction. This method uses Re-smearing function(RSF) to smear the lensed galaxy image and PSF image to make Re-smeared galaxy image(ReSMD) and Re-smeared PSF image(RePSF) which have the same ellipticity with that of the lensed galaxy.
The RSF must be a reasonable function which does not enhance the pixel noise on galaxy image so much. We tested this method using four types of PSF,  circular PSF, high elliptical PSF, double Gaussian PSF, PSF with pointing error.  We found that 
our new ERA  method is able to  correct PSF effect and obtained the ellipticity of galaxy with high precision. Thus new ERA method is able to analysis data with large PSF and/or PSF with high ellipticity without systematic error associated with PSF correction used in the previous approach.

For the practical observation, the systematic error caused by the pixel noises on smeared galaxies and on PSF images may not be ignored. We studied the effect of pixel noise and found that it makes overestimation of the order of 10\% for objects(SNR=20). 
Thus the pixel noise correction is needed for precise shear analysis in order to apply the weak lensing to an accurate determination of the cosmological parameters, in particularly the nature of dark energy property.  
We will study the pixel noise correction  in future works.

This work is supported in part by a Grant-in-Aid for Scientific Research from JSPS
(No.26400264 for T.F)


\end{document}